\documentclass[twocolumn, trackchanges]{aastex61}
\usepackage{natbib}
\usepackage{amsmath}

\submitjournal{ApJ}

\shorttitle{Residual Energy in Turbulence}
\shortauthors{Bowen et al.}

\begin{document}

\title{Impact of Residual Energy on Solar Wind Turbulent Spectra}

\correspondingauthor{Trevor A.~Bowen}
\email{tbowen@berkeley.edu}

\author{Trevor  A.~Bowen}
\affiliation{Space Sciences Laboratory,
University of California, Berkeley, CA}%
\affiliation{Physics Department,
University of California, Berkeley, CA}%

\author{Alfred Mallet}%
\affiliation{Space Sciences Center, 
University of New Hamphsire, Durham, NH}%

\affiliation{Space Sciences Laboratory,
University of California, Berkeley, CA}%

\author{John W. Bonnell}%
\affiliation{Space Sciences Laboratory,
University of California, Berkeley, CA}%

\author{Stuart D.~Bale}%
\affiliation{Space Sciences Laboratory,
University of California, Berkeley, CA}%
\affiliation{Physics Department,
University of California, Berkeley, CA}%

\newcommand{\bvec}[1]{{\ensuremath{\bf{#1}}}}
\newcommand{\e}[1]{\ensuremath{\times 10^{#1}}}
\def \wind {{\em Wind}}
\def \helios {{\em Helios}}

\def\dens[#1]{10$^{#1}$\hskip 1.5pt{cm$^{-3}$}}

\def\ex {\ensuremath {\hat{{\bvec{e}}}_1}}
\def\ey {\ensuremath {\hat{{\bvec{e}}}_2}}
\def\ez {\ensuremath {\hat{{\bvec{e}}}_3}}
\def\epar {\ensuremath {\hat{{\bvec{e}_\parallel}}}}

\def\cordnb {\ensuremath {C(\delta n, \delta B_\parallel)}}
\begin{abstract}
It is widely reported that the power spectra of magnetic field and velocity fluctuations in the solar wind have power law scalings with inertial-range spectral indices of -5/3 and -3/2 respectively. Studies of solar wind turbulence have repeatedly demonstrated the impact of discontinuities and coherent structures on the measured spectral index. Whether or not such discontinuities are self-generated by the turbulence or simply observations of advected structures from the inner heliosphere has been a matter of considerable debate. This work presents a statistical study of magnetic field and velocity spectral indices over 10 years of solar-wind observations; we find that anomalously steep magnetic spectra occur in magnetically dominated intervals with negative residual energy. However, this increase in negative residual energy has no noticeable impact on the spectral index of the velocity fluctuations, suggesting that these intervals with negative residual energy correspond to intermittent magnetic structures. We show statistically that the difference between magnetic and velocity spectral indices is a monotonic function of residual energy, consistent with previous work which suggests that intermittency in fluctuations causes spectral steepening. Additionally, a statistical analysis of cross helicity demonstrates that when the turbulence is balanced (low cross-helicity), the magnetic and velocity spectral indices are not equal, which suggests that our observations of negative residual energy and intermittent structures are related to non-linear turbulent interactions rather than the presence of advected pre-existing flux-tube structures.

\end{abstract}

\section{introduction}

%Below proton gyro-scales, the solar wind is often described as a conductive magnetohydrodynamic (MHD) fluid. 

Observations of power law spectral distributions of magnetic and kinetic energy in the solar wind, i.e. $E_\alpha \propto k^{\alpha}$, have led to the development of various theories of magnetohydrodynamic (MHD) turbulence. It is widely reported that magnetic energy in the inertial range follows a power law spectrum with $E_b \propto k_\perp^{-5/3}$, while kinetic energy follows a shallower power law spectrum of $E_v \propto{k_\perp^{-3/2}}$ \citep{Mangeneyetal2001,Salemetal2009,Podestaetal2007,Borovsky2012}. These spectral indices respectively support the theories of critically balanced turbulence and subsequent modifications accounting for the alignment between velocity and magnetic fluctuations \citep{GS95, Boldyrev2006}. The presence of $E \propto k_\perp^{-3/2}$ spectral distributions has been recovered in many subsequent numerical simulations \citep{PerezBoldyrev2009,Chandranetal2015,Malletetal2017}. %

%Observations of correlated v and b fluctuations have given rise to the idea of Alfv\'{e}nic turbulence in the solar wind. In the inertial range of solar wind turbulence, incompressible magnetohydrodynamic (MHD) turbulence is frequently constructed through the El\"{a}sser variables, with non-linear interactions occurring between z+ and z- fluctuations. However, the differences in spectral distributions of velocity and magnetic energies should preclude the existence of a self-similar turbulent cascade in the Elsasser variables.

It is known that discontinuities and intermittency in observations of turbulence affect measured spectral indices. \citet{RobertsGoldstein1987} identified large amplitude coherent and discontinuous structures resulting in steep $k^{-2}$ spectra. \citet{Lietal2011} showed that excluding intermittent current sheets from {\em{Ulysses}} magnetometer data led to the measurement of a  $E_b \propto k^{-3/2}$ scaling, rather than the typically reported $E_b \propto k^{-5/3}$ scaling. \cite{Borovsky2010} reconstructed the spectral distribution of magnetic field discontinuities of Advanced Composition Explorer (ACE) observations using a synthetic time-series, finding a $E_b \propto k_\perp^{-5/3}$ scaling. There are two dominant explanations for discontinuities and intermittency in the solar wind. The first suggests that discontinuities arise dynamically from the turbulent evolution of the plasma into current sheets \citep{Lietal2011,Salemetal2009,MininniPouquet2009,Matthaeusetal2015,Boldyrevetal2011,Changetal2004}. The second suggests that observations of discontinuities correspond to advected flux tube structures from the inner-heliosphere \citep{TuMarsch1993,Borovsky2008, Marianietal1973, Brunoetal2001,Brunoetal2007}.

It is also known that the solar wind contains statistically more magnetic than kinetic energy \citep{Bavassanoetal1998, Salemetal2009,Brunoetal1985, Robertsetal1987}. Various models of MHD turbulence under a range of physical conditions show the growth of negative residual energy, defined as $E_r=E_v-E_b$ \citep{MullerGrappin2005,Gogoberidzeetal2012,PerezBoldyrev2009,Boldyrevetal2011}. The normalized residual energy, 
\begin{equation}
\sigma_r =\frac{\langle v^2\rangle -\langle{b^2}\rangle}{\langle v^2\rangle +\langle{b^2}\rangle}=\frac{2 \langle \bvec{z}_+\cdot \bvec{z}_-\rangle}{\langle z_-^2\rangle + \langle z_+^2\rangle},
\end{equation}
is understood to quantify the relative dominance of magnetic or kinetic energy, or equivalently, the alignment between the Els\"{a}sser variables defined as $\bvec{z}_\pm= \bvec{v} \pm \bvec{b}/\sqrt{\mu_ 0\rho_0}$, where \bvec{v} and \bvec{b} are the fluctuating velocity and magnetic fields and $\rho_0$ is the mean mass density.

A power law spectrum for $E_r$ was derived by \cite{Grappin1983}, with ${E_r\propto k^{-2}}$ under the assumption of weak turbulence. \cite{MullerGrappin2005} have subsequently suggested $E_r\propto k^{-7/3}$ spectra for decaying isotropic turbulence and $E_r\propto k^{-2}$ scaling for forced anisotropic turbulence. \cite{Chenetal2013} used a statistical study of \wind{ } observations to explore connections between spectral index and residual energy, reporting a mean value of $\alpha_r$=-1.91 and a significant correlation between $\alpha_r$ and $\alpha_b$. In a study demonstrating scale invariance of normalized cross helicity 
\begin{equation}
\sigma_c=\frac{2\langle\delta\bvec{b}\cdot \delta\bvec{v}\rangle}{\langle v^2\rangle +\langle{b^2}\rangle}=\frac{\langle z_+^2\rangle -\langle z_-^2\rangle }{\langle z_-^2\rangle + \langle z_+^2\rangle},
\end{equation}
 \citet{PodestaBorovsky2010} reported $\alpha_r= -1.75$. Both studies demonstrate correlations between cross helicity and spectral indices for magnetic fields, velocity, as well as total energy.

The connection between cross helicity and residual energy is well established. \citet{Brunoetal2007} show that as fast solar wind evolves from 0.3 -1AU the distribution of {\em{Helios}} measurements moves from a highly cross helical (imbalanced) state to a state with low cross helicity (balanced) and high negative residual energy. \citet{Wicksetal2013a} studied the evolution of cross helicity and residual energy over injection and inertial scales, arguing that the mean angle between the Els\"{a}sser variables is scale dependent and maximized at the outer scale. \citet{Wicksetal2013b} show that observations of turbulence tend to be either strongly cross helical, or have strong residual energy.

In this Letter, we use 10 years of \wind{ }observations to study statistical connections between intermittency, magnetic discontinuities, residual energy, and spectral index. We demonstrate that discontinuous events are associated with magnetically dominated intervals with large negative residual energies. Intermittent discontinuities steepen the magnetic spectral index, but have little effect on the measured velocity spectra. Our observations are consistent with the generation of residual energy and intermittency through turbulence, and suggest a close link between residual energy and intermittency.
 
\section{Data}
We use observations from several instruments on the \wind{ } mission ranging 1996 January 1 through 2005 December 31: Magnetic Field Investigation (MFI) \cite{LeppingMFI}, Solar Wind Experiment (SWE) \cite{OgilvieSWE}, and Three Dimensional Plasma (3DP) experiment \cite{Lin3DP}. Data are separated into non-overlapping 1 hr intervals. Intervals are excluded if any of several conditions are met: {\em{Wind's}} geocentric distance is less than $35 R_E$, the average solar wind speed is $<$ 250 km/s, or if more than 5\% of of observations are missing from any one instrument. Linear interpolation is implemented across small data gaps when $< 5\%$  of an interval is missing. The resulting data consists of 39415 intervals of 1 hour.

The 3 s cadence 3DP ``on board" proton moment measurements are interpolated to the MFI time base. We separate velocity, and magnetic field measurements ($\bvec{v}$, and $\bvec{B}$) into mean and fluctuation quantities using time-averaged values, denoted as $\langle...\rangle.$ For example, the mean magnetic field, $\bvec{B_0},$ is determined by $\langle\bvec{B}\rangle=\bvec{B_0}$ with the fluctuation quantities as $\bvec{\delta B}=\bvec{B}-\bvec{B_0}$. We normalize the magnetic field to Alfv\'{e}n units using $\bvec{\delta b}=\bvec{\delta B}/\sqrt{\mu_ 0\rho_0}$ where $\rho_0$ is the mean mass density.

Each interval is characterized by energies associated with the velocity and magnetic field fluctuations  $E_b=\frac{1}{2} \langle \delta b^2 \rangle$ and $E_v=\frac{1}{2} \langle\delta v^2\rangle$, normalized cross helicity, \begin{equation}\sigma_c=\frac{2 \langle\delta\bvec{b}\cdot \delta\bvec{v}\rangle}{\langle{\delta b}^2\rangle+\langle{\delta v}^2\rangle},\end{equation} and normalized residual energy \begin{equation}\sigma_r =\frac{E_v -E_b}{E_v +E_b}.\end{equation}

A minimum variance analysis (MVA) is performed on $\delta\bvec{v}$ and $\delta\bvec{b}$ to decompose each interval into eigenvectors corresponding to directions of minimum, maximum, and intermediate variance \citep{SonnerupCahill1963}. Intervals with maximum energy is largely distributed along a single direction, i.e. if ${\lambda_b^{max}} \approx E_b$, may indicate the presence of strong discontinuities or a linear polarization to the fluctuations \citep{Brunoetal2001}.

Intermittency in the magnetic field is often associated with current sheets \cite{Matthaeusetal2015,VeltriMangeney1999,MininniPouquet2009, Malletetal2016}. Using Ampere's law \begin{equation}\nabla \times \bvec{B} =\mu_0\bvec{J}\end{equation} and invoking the Taylor hypothesis, $\frac{\partial}{\partial t} \sim \bvec{V}\cdot\nabla$ allows the time derivative of magnetic field observations in the spacecraft frame to be used as a proxy for current \citep{Podesta2017}. Because single spacecraft observations constrain spatial derivatives to the bulk solar wind flow direction, the full curl cannot be computed. To estimate the magnitude of currents we implement the reduced curl \begin{equation} \nabla_x \times \bvec{B} = -\frac{\partial}{\partial x} B_z\hat{y} +\frac{\partial}{\partial x} B_y\hat{z},\end{equation} where the solar wind flow is along $\hat{x}$. Applying the Taylor hypothesis gives an estimate of the current magnitude, \begin{equation}J=\frac{1}{\mu_0V_{sw}} \sqrt{\left(\frac{\partial B_y}{\partial t}\right)^2 +\left(\frac{\partial B_z}{\partial t}\right)^2}.\end{equation} A reduced estimate for the vorticity magnitude $\bvec{\omega} =\nabla \times \bvec{v}$ is similarly computed.

Intermittency is frequently quantified using the kurtosis, \begin{equation}\kappa_x=\frac{\langle x^4\rangle}{\langle x^2\rangle^2}\end{equation}\citep{Brunoetal2001,Brunoetal2003,Salemetal2009,Mangeneyetal2001,VeltriMangeney1999,Frisch1995,Matthaeusetal2015}. Gaussian distribution have  $\kappa=3$, with $\kappa > 3$ indicating heavy tailed, non-Gaussian statistics. As a simple statistic to quantify intermittency in the magnetic and velocity fluctuations we measure the kurtosis of the reduced curl estimations of the current and vorticity, $\kappa_{J}$ and $\kappa_{\omega}$ for each interval, subtracting 3 to compare with Gaussian statistics. 

\section{Spectral Fitting}
Trace spectral indices for the magnetic and velocity fluctuations in the inertial turbulent range are estimated by performing a linear least squares fit of the power spectra to a line in logarithmic space. Power spectra for $\bvec{b}(t)$ and $\bvec{v}(t)$ are estimated with a fast-Fourier transform. The trace power spectra, $\tilde{E_b}$ and $\tilde{E_v}$ are calculated as the sum power spectra from each direction axis. 

To prevent overlapping with injection scales, our fits only consider frequencies above $\sim0.277$ Hz  (6 minutes). To avoid spectral steepening associated with the dissipative scales at high frequencies, we only consider the subsequent 190 frequency bins (up to 0.0555 Hz, or 18 seconds). The trace spectra are linearly interpolated to an abscissa of 50 logarithmically spaced frequencies (linearly spaced in the logarithmic domain) between 0.277-0.0555 Hz. The power spectra is estimated using a linear least square fit of the interpolated spectra and frequencies in log-log space, with the slope of the best fit line giving the spectral index \citep{Podesta2016, Chenetal2013}. The spectral index of the trace residual energy is calculated from fitting $|\tilde{E}_r|=|\tilde{E_v}-\tilde{E_b}|$ with the same interpolation and least square fitting scheme. Additionally, the high frequency limit helps to minimize flattening effects due Gaussian noise in low amplitude 3DP velocity measurements; though the range of our spectral fits extends to slightly higher frequencies than what previous authors have used, we find good agreement with their estimates for mean values of $\alpha_v$ and $\alpha_b$ \citep{Chenetal2013, Wicksetal2013a,PodestaBorovsky2010}. 

Uncertainty of our estimated spectral indices is found through propagation of error \citep{NumC}. The variance associated with single FFT estimation of spectral density is equal to the power spectral density itself. Typically, variance is reduced through averaging over an ensemble of spectra, or windowing the autocorrelation function of a time-series. Here we derive the uncertainty in spectral index associated with least squares fitting of a single FFT estimation of spectral density. For spectral density $S_i$ where index $i$ refers to a given frequency bin, $f_i,$ \cite{stoica} give the variance of the spectral density as \begin{equation}\text{Var}[{S_i}]=\sigma^2_i\approx S_i^2.\end{equation}
Propagating the variance $\sigma^2_i$ to the logarithm of the spectral density $\text{log}_{10}(S_i)$ gives \begin{equation}\text{Var}[\text{log}_{10}S_i]=\sigma^2_{L}=\left(\frac{1}{\text{ln}10}\right)^2
\frac{\sigma_i^2}{S_i^2}\approx {0.19}.\end{equation} The scaling of $\text{Var}[\tilde{S_i}]=S_i$ leads to constant variance in the estimation of the logarithm of spectral density.

For a power law spectra $S_i=\beta f_i^{\alpha}$ minimizing \begin{equation}\chi^2 =\sum_{i=0}^{N-1} \left(\frac{y_i -\beta -\alpha x_i}{\sigma_L}\right)^2,\end{equation} where $y_i=\text{log}_{10}S_i$ and $x_i=\text{log}_{10}f_i $, with respect to $\alpha$ and $\beta$ gives the least square best fits for the spectral index and scaling amplitude. Following \citet{NumC} for propagation of errors gives the uncertainty in $\alpha$ as \begin{equation}\sigma^2_\alpha = \sum_i\left(\frac{\partial{\alpha}}{\partial{y_i}}\right)^2\sigma_L^2=\sigma_L^2 \frac{\sum x_i^2}{N\sum x_i^2 -(\sum x_i)^2}.\end{equation} The uncertainty in the estimated spectral index, a function of $\sigma_L^2$ and the uniformly used frequency abscissa, is constant for each interval with $\sigma_\alpha =\pm 0.16$.
\section{Results}

\begin{figure*}
\centering
\includegraphics[width=12.7 cm]{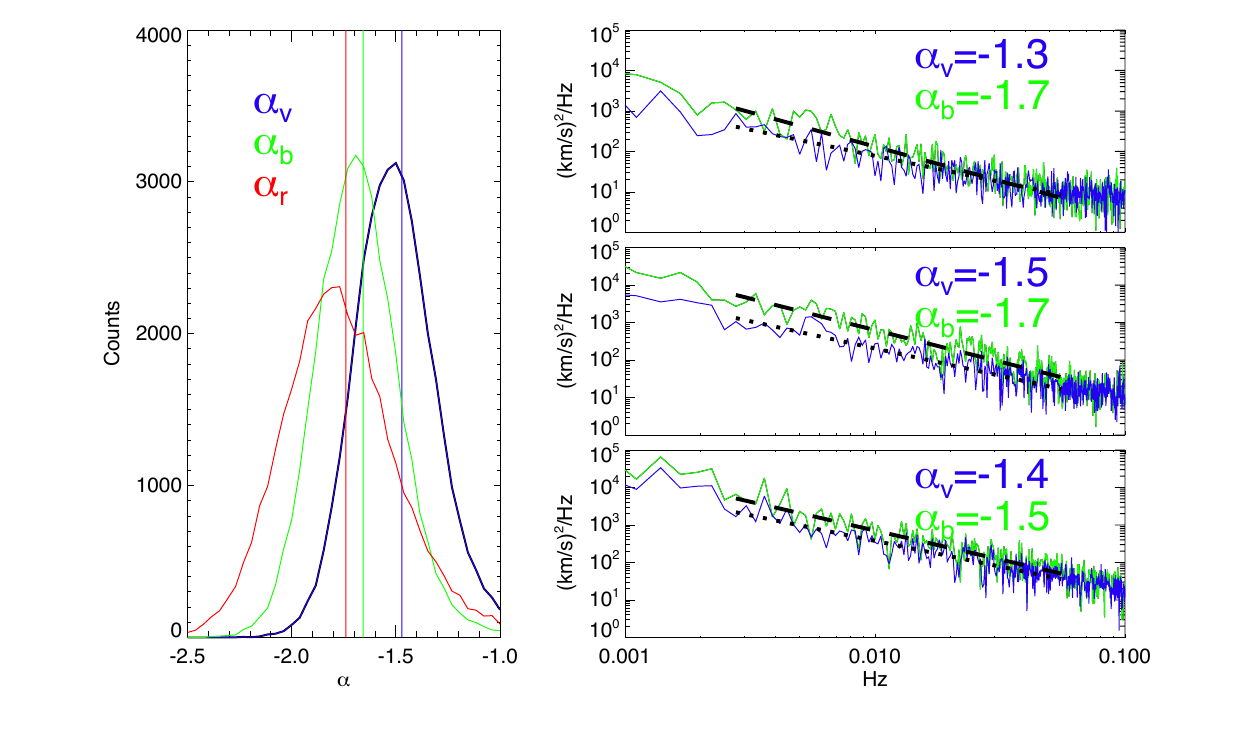}
\caption{(Left) Distribution of fits power-law indices for magnetic field (green), velocity (blue), and residual energy (red) spectra, mean values are shown with vertical lines. (Right) Examples of measured magnetic field (in Alfv\'{e}n units) and velocity fluctuation spectra. Fits for the magnetic and velocity spectra are shown respectively as black dashed and dotted lines.}
\label{fig:fig1}
\end{figure*}

\begin{figure*}
\centering
\includegraphics[width=12.7 cm]{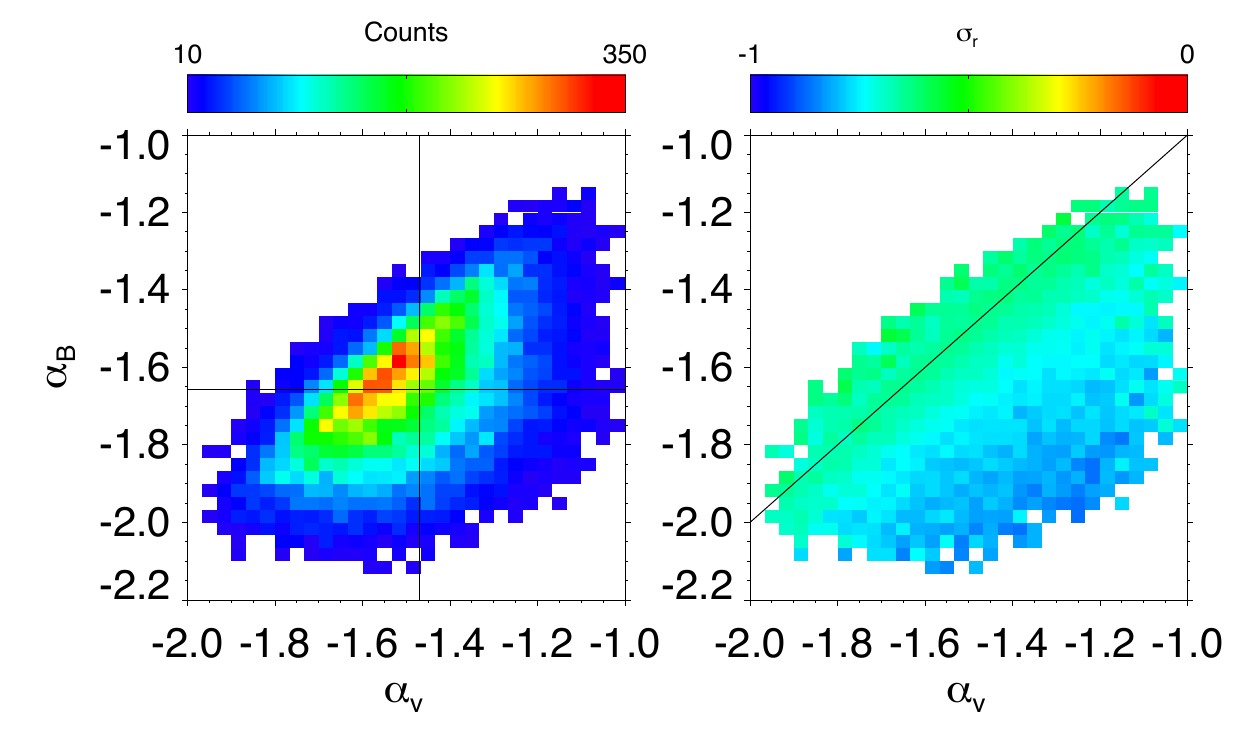}
\caption{(Left) Joint distribution of the fitted spectral indices for magnetic field, $\alpha_b$, and velocity, $\alpha_v$, fluctuations in the inertial range. The black lines show the mean values of $\alpha_v \sim-3/2$ and $\alpha_b\sim -5/3.$ (Right) The distribution of $\alpha_v$ and $\alpha_b$ colored by the mean residual energy in each bin. The black line shows  $\alpha_v =\alpha_b$.  Deviations from $\alpha_v \approx \alpha_b$ lead to an increase in negative residual energy ($E_b>E_v$). }
\label{fig:fig2}
\end{figure*}

Figure 1 shows the probability distributions of $\alpha_{b}$, $\alpha_{v}$, and $\alpha_r$, with respective means of -1.66,-1.47, and -1.73. Our fits for the the velocity and magnetic energy spectra agree with spectral indices given in previous studies \citep{Mangeneyetal2001,Salemetal2009,Podestaetal2007,Borovsky2012}. The mean value of $\alpha_r$ slightly shallower than observations in  \citet{Chenetal2013} but is consistent with \citet{PodestaBorovsky2010}. Our observations of $\alpha_r$ are also shallower than predictions of various models of MHD turbulence \citep{Grappin1983,MullerGrappin2005,Gogoberidzeetal2012}; however, these models are conducted using assumptions which are not satisfied by solar wind turbulence, e.g. weak turbulence, isotropy, and quasi-normal closure. Deviation in our measurements of $\alpha_r$ from \citet{Chenetal2013} likely occur due to differences in the fitting technique and normalization of the magnetic field. Our work directly fits $\tilde{E}_r$ as the difference in observed velocity and magnetic spectra, and implements MHD normalization of the magnetic field. \cite{Chenetal2013} use fitted spectra to calculate $\alpha_r$ and implement a kinetic normalization of the magnetic field. The right panels of Figure 1 show examples of $\tilde{E_b}$, $\tilde{E_v}$, and $\tilde{E_r}$ with our fits.

\begin{figure*}[h]
\centering
%\plotone{figs/parametric_ms_fig2.eps}
\includegraphics[width=12.7cm]{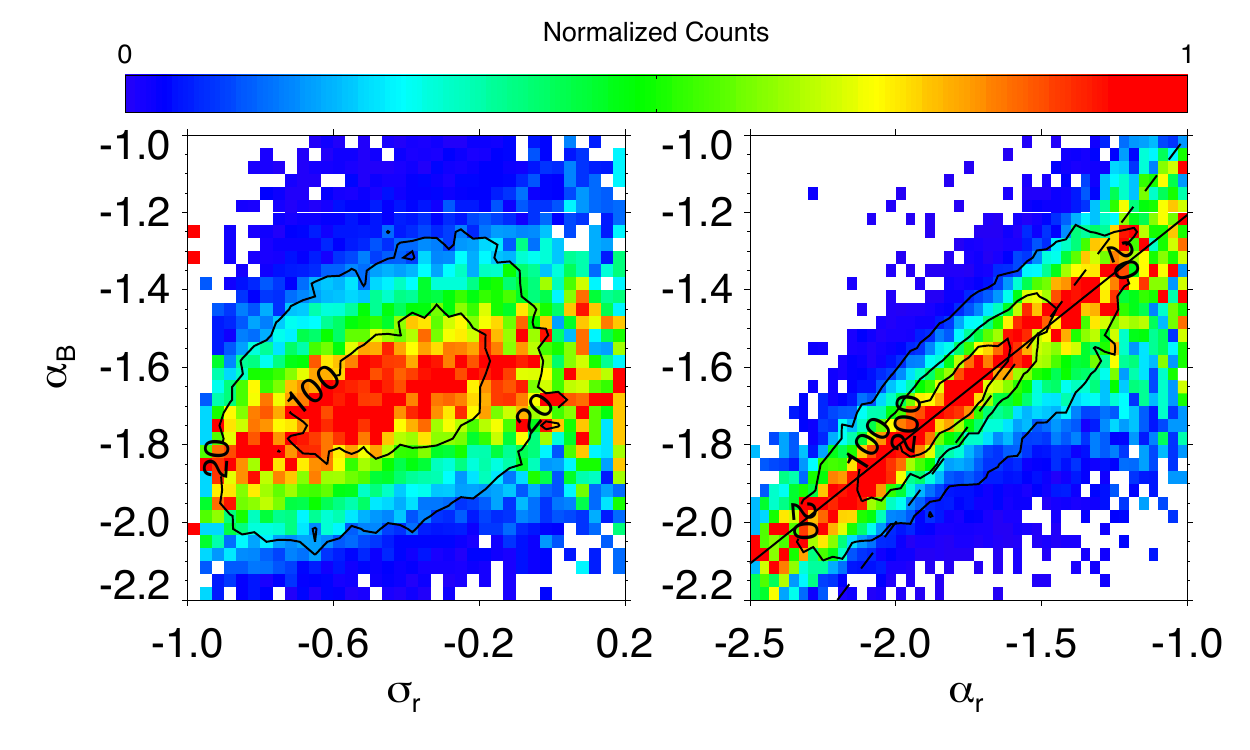}
\caption{(Left) Joint distribution of normalized residual energy $\sigma_r$ and spectral index of the inertial range magnetic fluctuations $\alpha_{b}$. Data is column normalized to the maximum number of counts in each $\sigma_r$  bin. (Right) Joint distributions of fitted power law spectral indices of residual energy spectra $\alpha_r$ and inertial range magnetic field fluctuations $\alpha_{b}$.  The solid black line shows the least square linear fit to the data with correlation 0.77 and slope of 0.56. The dashed line shows $\alpha_r=\alpha_b$. Contours in either panel show 20, 100, and 200 level counts of the joint distributions.}
\label{fig:fig3}
\end{figure*}

The left panel of Figure 2 shows the joint distribution of $\alpha_b$ and $\alpha_v$. The mean value of the velocity spectral index is $\alpha_v=-3/2$. A tendency for $\alpha_b < \alpha_v$ is evident in the distribution. The right panel of Figure 2 shows the joint distribution of magnetic and velocity indices colored by the mean value of $\sigma_r$. The statistical preference for negative residual energy is clearly evident in our observations. The residual energy becomes more negative as the spectral indices $\alpha_v$ and $\alpha_b$ diverge, i.e. as the magnetic spectral index steepens. Particularly interesting is the consistent level of residual energy along the line $\alpha_v =\alpha_b$ throughout the range of observations.

The left panel of Figure 3 shows the joint distribution of $\sigma_r$ and $\alpha_b$. The secular trend suggests that the residual energy plays a significant role in setting the spectral index of the magnetic field. Specifically, it is evident that magnetically dominated intervals, with $\sigma_r \approx -1,$ exhibit steeper magnetic spectra. The right panel of Figure 3 shows the joint distribution of  $\alpha_r$ and $\alpha_b$, these variables are highly correlated with a Pearson correlation value of 0.78. A linear best fit gives $\alpha_b \propto 0.56 \alpha_r$. These results imply that spectral indices of the magnetic fluctuations and residual energy are largely determined by the average residual energy over each interval.

\begin{figure*}[h]
\centering
%\plotone{figs/parametric_ms_fig3.eps}
\includegraphics[width=12.7cm]{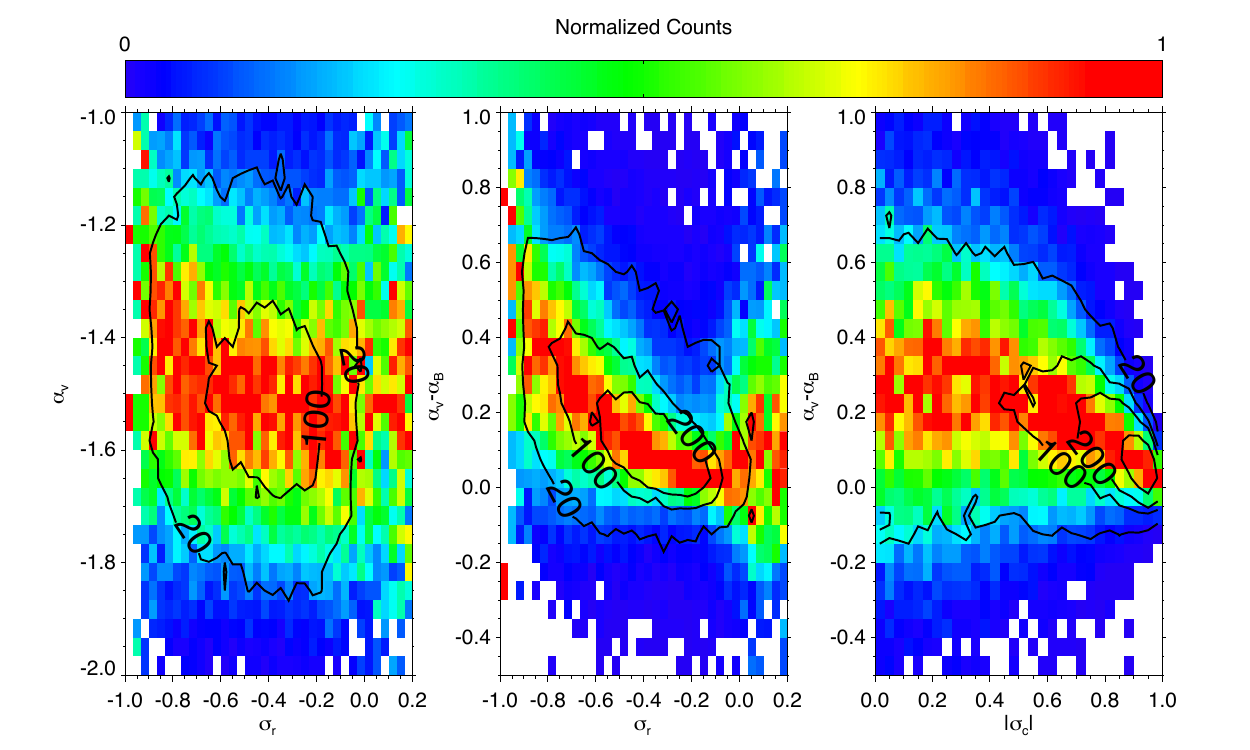}
\caption{(Left) Joint distribution of fit velocity spectral index $\alpha_v$ with root-mean-square residual energy $\sigma_r$. The distribution is column normalized to the maximum occurrence of $\alpha_v$ for each value of $\sigma_r$. (Center) Joint distribution of the difference in velocity and magnetic field spectral indices, $\alpha_v -\alpha_b$, with $\sigma_r$. (Right) Joint distribution of the difference in velocity and magnetic field spectral indices, $\alpha_v -\alpha_b$, with $\sigma_c$. Data is column normalized to the maximum number of counts of $\alpha_v -\alpha_b$ in each bin of $\sigma_c$. Contours in all three plots show 20, 100, and 200 count levels of the distributions.}
\label{fig:fig4}
\end{figure*}

The left panel of Figure 4 shows the joint distribution of $\sigma_r$ and $\alpha_v$. Unlike the spectral index of the magnetic fluctuations, $\alpha_v$ exhibits little dependence on $\sigma_r$, suggesting that residual energy is mostly determined by magnetic fluctuations. The middle panel of Figure 4 shows the difference between the magnetic and velocity spectral indices as a function of residual energy. As the residual energy increases, i.e. the plasma becomes less magnetically dominated, the spectral index of the magnetic field approaches that of the velocity spectra.
%, consistent with the idea of magnetic intermittency or discontinuous features affecting the spectral index \citep{Salemetal2009,Lietal2011}. 

The right panel of Figure 4 shows the joint distribution of the difference between magnetic and velocity spectral indices, $\alpha_v-\alpha_b$, and cross helicity, $\sigma_c$, suggesting that high cross helicity measurements occur only when $\alpha_v =\alpha_b$. A geometrical consideration of cross helicity and residual energy gives the constraint of $\sigma_c^2+\sigma_r^2 <1$ \citep{Wicksetal2013b}.  Clearly, the decrease of $|\sigma_c|$ with large negative $\sigma_r$ and $|\alpha_b| > |\alpha_v|$ is inevitable. However, there is no such geometric argument which demands balanced turbulence (i.e. $|\sigma_c| <1$) to coincide with unequal spectral indices such that $\alpha_b \ne \alpha_v.$ The observations in Figure 4 (Right), in which the joint distribution of $\alpha_v-\alpha_b$ is conditioned on $\sigma_c$ suggests that balanced turbulence (i.e. $|\sigma_c| < 1$) coincides with $|\alpha_b| > |\alpha_v|$. There is no {\em{a priori}} reason that we expect balanced turbulence (i.e. $|\sigma_c| <1$) to have different spectral indices, $\alpha_b \ne \alpha_v.$ This result is consistent with the generation of negative residual energy through turbulence, i.e. that non-linear interactions between the Els\"{a}sser variables lead to the growth of intermittent structures with negative residual energy, since at fixed total energy the nonlinear interaction term $ \bvec{z_\pm}\cdot \nabla \bvec{z_\mp}$ are stronger when $\sigma_c =0.$

%
% Otherwise, we would expect the presence of balanced, i.e. low cross helictity observations with equal spectral slopes--i.e. the plasma evolves from a high cross helicity, low residual energy state, to a balanced state with higher residual energy.

Using the MVA analysis, the value ${\lambda^{max}}$ corresponding the fraction of energy associated with the maximum variance direction, is calculated for both the magnetic and velocity fluctuations. The top panels of Figure 5 show the joint distributions of ${\lambda^{max}_b}$ (Left) and ${\lambda^{max}_v}$ (Right) with the residual energy. There is a strong dependence of ${\lambda^{max}_b}$ on the residual energy which is not observed for ${\lambda^{max}_v}$. The suggests that large negative residual energy occurs as the result of discontinuous/coherent structures in the magnetic field. In fact, the most negative values of residual energy seem to demonstrate the smallest values of ${\lambda^{max}_v}$, which suggest more isotropic velocity fluctuations; however, this could be due to sampling bias towards very low amplitude velocity fluctuations subject to noise.

To further connect the negative residual energy with magnetic intermittency, we examine the kurtosis of the reduced curl estimates for current and vorticity, $\kappa_{J}$ and $\kappa_{\omega}$, as proxies for intermittent features. The bottom panels of Figure 5 show the joint distribution of the residual energy and $\kappa_{J}$. A decrease in $\kappa_{J}$ is observed with increasing residual energy, suggesting that the negative residual energy is caused by magnetic discontinuities with associated bursty currents. At low residual energy the velocity fluctuations appear more Gaussian, which may indicate very low amplitude velocity fluctuations possibly subject to noise. Regardless, we uniformly observe $\kappa_{\omega} <\kappa_{J},$ suggesting less intermittency in the velocity fluctuations.

 \begin{figure*}[h]
 \centering
\includegraphics[width=18cm]{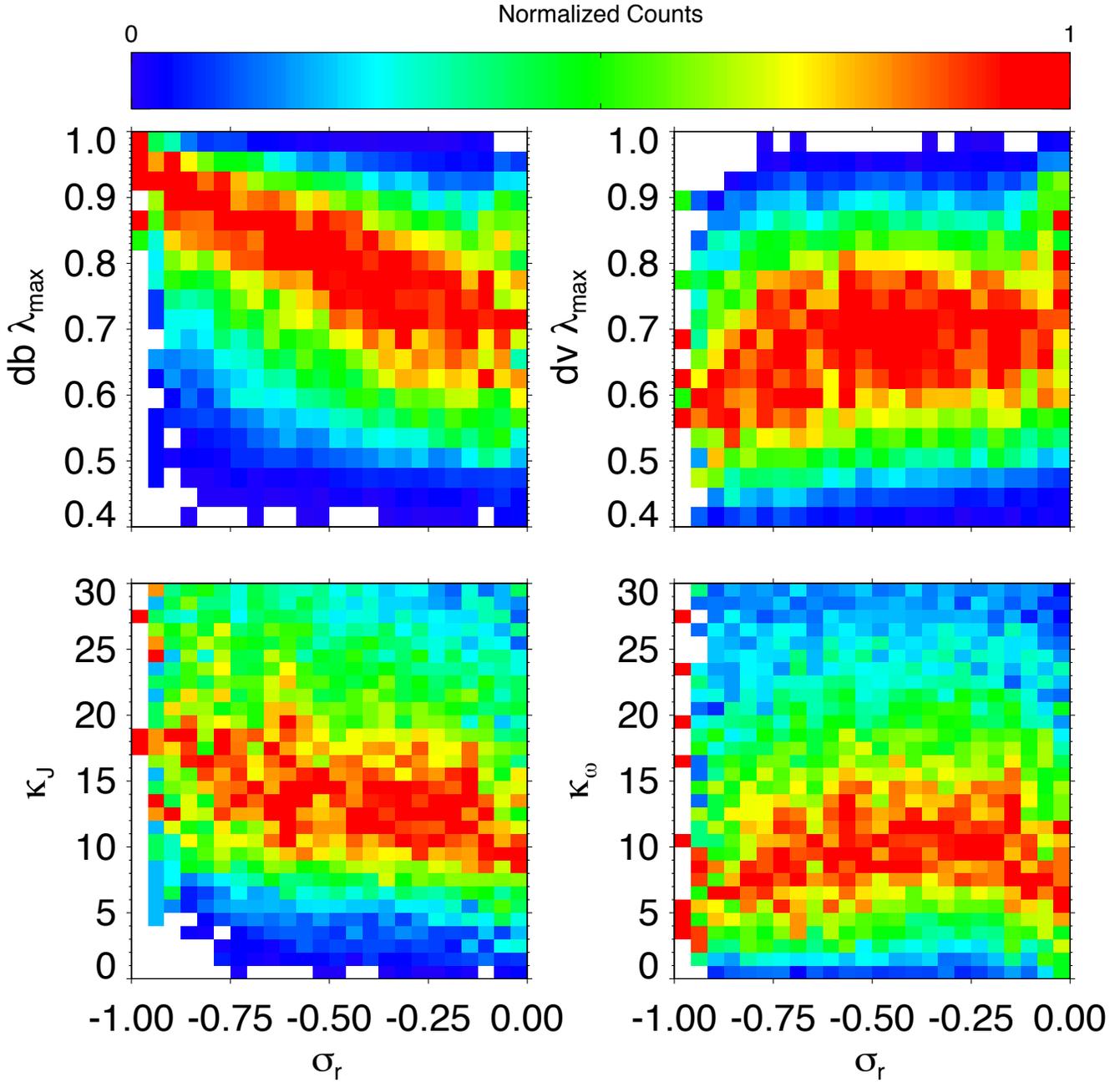}
\caption{(Top Left) Joint distribution of maximum normalized eigenvalue of magnetic field fluctuations, $\lambda_b^{max},$ with $\sigma_r.$ Large negative residual energy corresponds with large $\lambda_{b}^max$. The joint distribution of $\lambda_v^{max}$ with the residual energy (Top Right) does not show a dependence on the residual energy.  Joint distributions of the kurtosis of reduced current (Bottom Left) and vorticity (Bottom Right) with residual energy show that negative residual energy corresponds to intermittent currents with no associated signature in the velocity fluctuations.}
\label{fig:fig5}
\end{figure*}
 \section{Discussion}

%
%\begin{enumerate}
%\item{Magnetic spectral lopes are ordered by residual energy, Fig 1 left and right}
%
%\item{Velocity spectral slopes are largely insensitive to residual energy, Fig 3 Left}
%
%\item{The velocity spectra with the rms residual energy, effectively sets the trace of the magnetic spectra, Fig 3 Left and Center}
%\item{It is likely the excess of magnetic energy relative to kinetic energy generates the average b$\sim$-5/3 vs v$\sim-3/2$ slopes. known in Li et al as intermittency. But here shown more generally for all spectral slopes }
%\item{The velocity spectral slope is a random variable with some mean and dispersion, Fig 3 Left}
%\item{The difference in b and v spectral indices depends on the amount of residual energy. Fig 3 Center}
%\item{Cross helicity becomes large when spectral slopes are equal, suggesting that fluctuations in excess magnetic energy is not correlated with dv fluctuations, Fig3 Right}
%\item{Excess magnetic energy corresponds to large eigenvalue decompositions in the magnetic field which are not present in the velocity field decomposition. These large excess magnetic energy intervals are observed to have large currents.}
%\end{enumerate}
%
%
%

%%
%%		Figure 1 - Hc and XCC vs theta_sm
%%
 
 Many authors recover $k^{-3/2}$ spectra in simulations \citep{MaronGoldreich2001,MullerGrappin2005,PerezBoldyrev2009,Malletetal2016}. This scaling is in agreement with analytic predictions of strong, three dimensional, anisotropic, turbulence appropriate for the solar wind \citep{Boldyrev2006,Chandranetal2015,MalletSchekochihin2017,BrunoCarbone2013}. Our observations here suggest that the observed difference between the spectral indices of velocity and magnetic field turbulent fluctuations occurs due to the presence of negative residual energy in the form of intermittent current sheets.  When the magnetic and velocity energies are in equipartition, the spectral slope of the magnetic fluctuations approaches the velocity spectral index. The velocity spectral index is insensitive to the residual energy with a mean value of $\alpha_v=-3/2$. This picture is consistent with the numerical model of \cite{MininniPouquet2009}, which demonstrates the formation of thin current sheets in the magnetic field through decaying turbulence leading to enhanced intermittency and steepening of magnetic spectral index of $\alpha_b=-5/3$. In this interpretation, the magnetic fluctuations form thinner structures than the velocity fluctuations, which then dissipate energy more quickly. This picture does not address the collisionless and kinetic nature of dissipation in the solar wind; a full explanation requires a more complex account of the physical mechanisms of dissipation.

Our results are congruent with \cite{Lietal2011} who interpreted their results as indicative of flux-tube crossings. However, for several reasons, we believe our results support the idea of intermittency through turbulence rather than observations of advected flux tubes. First, we have identified the presence of intermittent events in the magnetic field contributing to negative residual energy which have no accompanying signature in the velocity fluctuations. Observed intervals with intermittent signatures present in both velocity and magnetic fluctuations are likely contained along the $\alpha_v=\alpha_b$ line, where steepening may occur in both the magnetic and velocity spectra. If observations of flux tube crossings are present in the dataset, they likely exist in this region. Additionally, our results agree with \cite{Salemetal2009} who note that the high kurtosis distributions which affect measurements of spectral indices occur at lower fluctuation amplitudes in the magnetic field than in the velocity measurements. This again suggests the presence of intermittent magnetic fluctuations with no velocity component.

The joint distribution of the difference of spectral slopes, $\alpha_v -\alpha_b$, and the cross helicity $\sigma_c$, suggests that intervals of balanced turbulence preferentially occur with $\alpha_v \ne\alpha_b.$ Though unequal spectral slopes, associated with non-equipartitioned $E_v$ and $E_b$, geometrically preclude the observation of imbalanced fluctuations with $|\sigma_c| \sim 1$, the observations in Figure 4 (Right) suggest the stronger statement that balanced turbulence occurs only with unequal spectral indices. We interpret the lack of balanced turbulence with $\alpha_v =\alpha_b$ as evidence for the generation of residual energy through non-linear turbulent interactions \citep{Boldyrevetal2011}. The observation that solar wind turbulence is either highly imbalanced, $|\sigma_c |=1$ or highly anti-aligned $\sigma_r =-1$ has been noted by previous authors \citep{Wicksetal2013b,Brunoetal2007}; however, observations of low cross helicity directly corresponding directly to deviations in turbulent spectral indices suggests that the residual energy is closely connected with non-linear turbulent interactions.

The quantification of the variance in spectral density estimates demonstrates that our fit spectral indices are accurate to ~10\%. The implementation of this variance estimate will help constrain observations made of the inner heliosphere by the FIELDS instrument on the Parker Solar Probe \citep{FIELDS}. Additionally, a quantitative characterization of spectral index variance may prove useful in further studies of \wind{} observations, e.g. determining the nature compressive fluctuations in the solar wind \citep{Bowenetal2018}.

\section{Acknowledgements}
A. Mallet would like to acknowledge useful conversation with B.D.G. Chandran and A.A. Schekochihin. T.A.B. was supported by NASA Earth and Space Science Fellowship NNX16AT22H. A. Mallet was supported by NSF grant AGS-1624501. Wind/3DP data analysis at UC Berkeley is supported in part by NASA grant NNX16AP95G.

\end{document}